\documentclass[final,authoryear,5p,times,twocolumn]{elsarticle}

\usepackage{amssymb}

\journal{Neuroimage}

\usepackage{graphicx}
\usepackage{color}
\usepackage{version}
\usepackage[colorlinks]{hyperref}
\usepackage[normalem]{ulem}
\usepackage[dvipsnames]{xcolor}

\newcommand\eg[0]      {e.g.}

\newcommand\cf[0]      {\textit{cf.}}

\newcommand\copysim    {\mathrm{\scalebox{0.65}{\textsf{\copyright}}}}
\newcommand\figref[1]  {Fig.~\ref{#1}}

\begin{document}

\begin{frontmatter}



\title{Solved problems and remaining challenges for Granger causality analysis in neuroscience: A response to Stokes and Purdon (2017)}


\author{Lionel Barnett, Adam B. Barrett, Anil K. Seth*}

\address{Sackler Centre for Consciousness Science and Department of Informatics, \\
University of Sussex, Brighton BN1 9QJ, UK\\
*a.k.seth@sussex.ac.uk (correspondence)}

\begin{abstract}
Granger-Geweke causality (GGC) is a powerful and popular method for identifying directed functional (`causal') connectivity in neuroscience. In a recent paper, \cite{StokesPurdon:2017} raise several concerns about its use. They make two primary claims: (1) that GGC estimates may be severely biased or of high variance, and (2) that GGC fails to reveal the full structural/causal mechanisms of a system. However, these claims rest, respectively, on an incomplete evaluation of the literature, and a misconception about what GGC can be said to measure. Here we explain how existing approaches [as implemented, for example, in our popular MVGC software \citep{Barnett:mvgc:2012}] resolve the first issue, and discuss the frequently-misunderstood distinction between \emph{functional} and \emph{effective} neural connectivity which underlies Stokes and Purdon's second claim.
\end{abstract}


\end{frontmatter}



Granger-Geweke causality (GGC) is a powerful analysis method for inferring directed functional (`causal') connectivity from time-series data, which has become increasingly popular in a variety of neuroimaging contexts \citep{Seth:jneuro:2015}.  GGC operationalises a statistical, predictive notion of causality in which causes precede, and help predict their effects. When implemented using autoregressive modelling, GGC can be computed in both time and frequency domains, in both bivariate and multivariate (conditional) formulations. Despite its popularity and power, the use of GGC in neuroscience and neuroimaging has remained controversial. In a recent paper, \cite{StokesPurdon:2017} raise two primary concerns: (1) that GGC estimates may be severely biased or of high variance, and (2) that GGC fails to reveal the full structural/causal mechanisms of a system. Here, we explain why these concerns are misplaced.

Regarding the first claim, \cite{StokesPurdon:2017} describe how bias and variance in GGC estimation arise from the use of separate, independent, full and reduced regressions. However, this problem has long been recognised \citep{Chen:2006, Barnett:mvgc:2014} and, moreover, has already been solved by methods which derive GGC from a single full regression\footnote{We note here that the ``partition matrix'' solution proposed by \cite{Chen:2006} is incorrect; see, \eg, \cite{Solo:2016}.}. These methods essentially extract reduced model parameters from the full model via factorisation of the spectral density matrix. Well-documented approaches include Wilson's frequency-domain algorithm \citep{Dhamala:2008a}, Whittle's time-domain algorithm \citep{Barnett:mvgc:2014}, and a state-space approach which devolves to solution of a discrete-time algebraic Riccati equation \citep{Barnett:ssgc:2015, Solo:2016}. Thus, the source of bias and variance discussed in \cite{StokesPurdon:2017} has already been resolved.

\begin{figure}
\begin{center}
\includegraphics{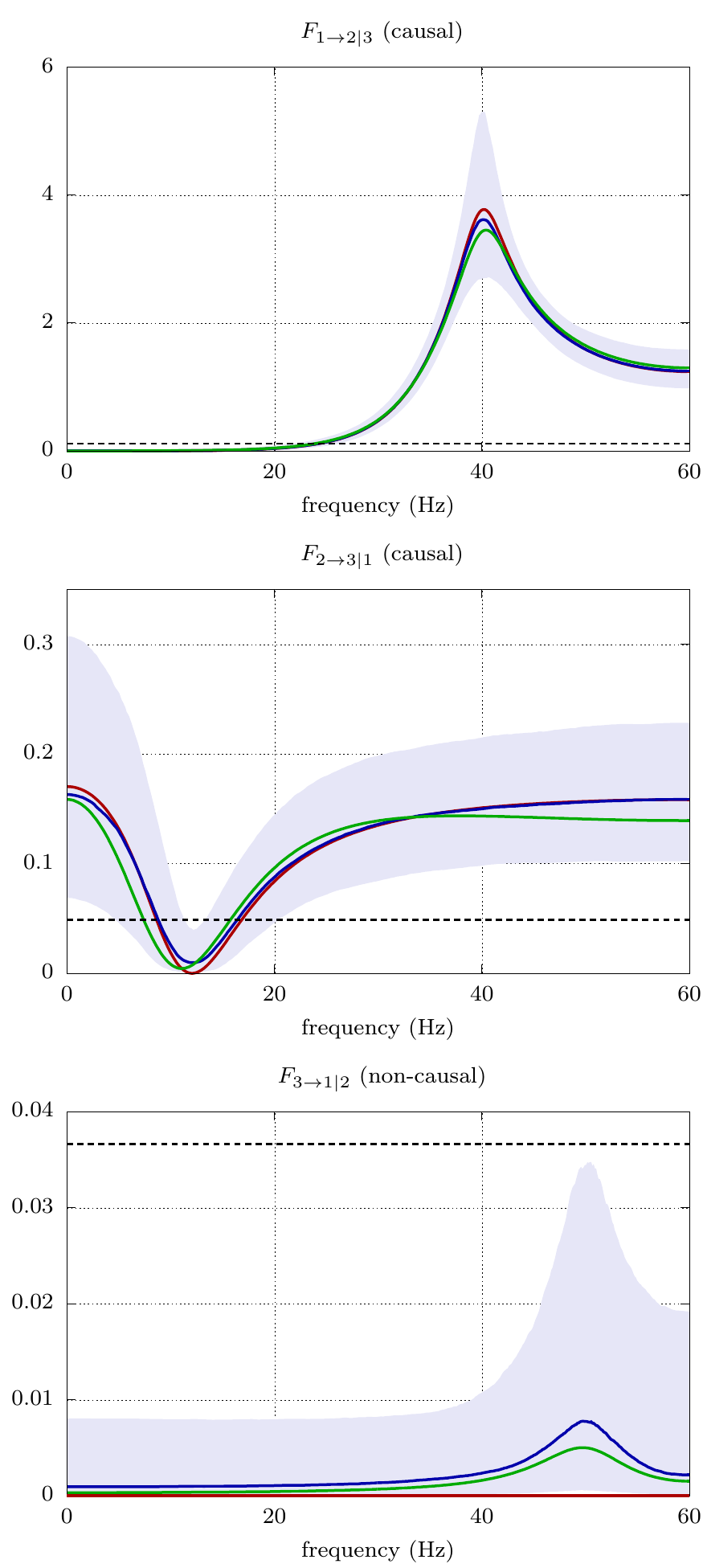}
\caption{Granger-Geweke frequency-domain causalities estimated by the single-regression state-space method \citep{Barnett:ssgc:2015, Solo:2016} for the $3$-node VAR model in \cite{StokesPurdon:2017}, (Example 1, \cf~Fig.~2). The true model order of $3$ was used for the (single, full-model) VAR estimates. Plots are based on $10,000$ time series realisations of $500$ observations: red lines plot the exact causality for the model and blue lines sample estimate medians. The shaded areas indicate 90\% central confidence intervals, while the green lines plot. The dashed horizontal lines indicate critical thresholds over all frequencies [see \cite{StokesPurdon:2017}, Supporting Information, S9] at $95\%$ significance, derived from simulation of the corresponding null model.} \label{fig:sgc}
\end{center}
\end{figure}

This is clearly illustrated in \figref{fig:sgc}, where we plot estimated frequency-domain GGC for the $3$-node VAR model in \cite{StokesPurdon:2017}, Example 1, using the single-regression state-space method \citep{Barnett:ssgc:2015, Solo:2016}. We remark that identical results are obtained using the time-domain spectral factorisation method of \cite{Barnett:mvgc:2014}, as implemented in the current (v1.0, 2012) release of the associated MVGC Matlab$^\copysim$ software package \citep{Barnett:mvgc:2012}. \figref{fig:sgc} may be directly compared with Fig.~2 in \cite{StokesPurdon:2017}; we see clearly that all estimates are strictly non-negative, and that exaggerated bias and variance associated with the dual-regression approach are absent. Therefore,  \cite{StokesPurdon:2017} are in error when they state that ``Barnett and Seth [\ldots] have proposed fitting the reduced model and using it to directly compute the spectral components \ldots''. This is important to note because our MVGC toolbox has been widely adopted within the community, with $> 3,500$ downloads and a significant number of high-impact research publications using the method \citep[\eg,][]{YellinEtal:2015,BruneauEtal:2015,PlaceEtal:2016,SchmittEtal:2017,AaronEtal:2017}. Thus, we can reassure users of the toolbox that problems of bias and variance as described by \cite{StokesPurdon:2017} do not apply.

\begin{figure}
\includegraphics{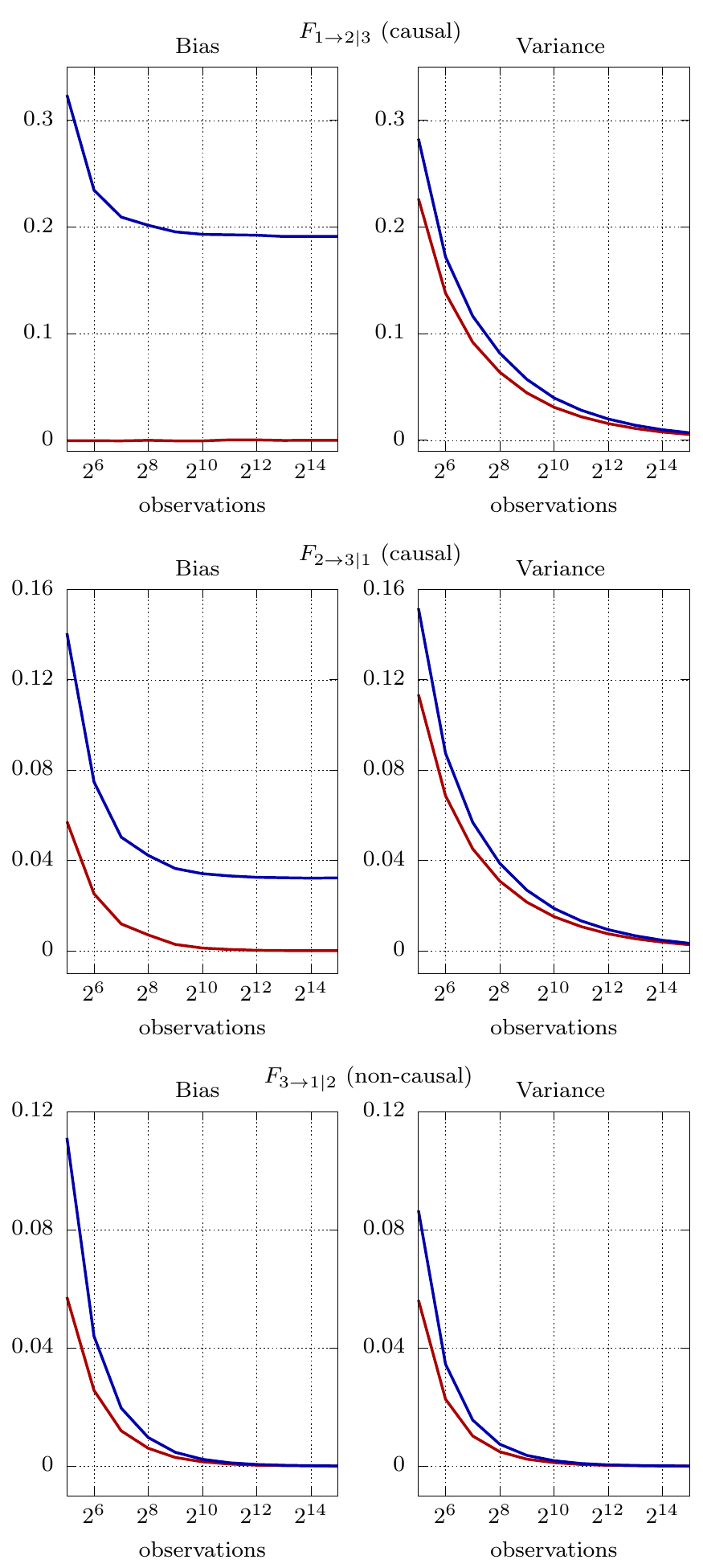}
\caption{Granger-Geweke time-domain causality bias (left column) and variance (right column) for estimation by the single-regression state-space method (red lines) and dual-regression method (blue lines), plotted aganst time series length, for the example $3$-node VAR model in \cite{StokesPurdon:2017}. Bias is measured as the difference between the sample median and true causality, while variance is measured as the mean absolute deviation of the sample causality (we use non-parametric measures, as the GGC sample estimators are non-negative, non-Gaussian, and potentially highly skewed). The true model order of $3$ was used for all VAR estimates. Plots are based on $10,000$ time series realisations for each number of observations. } \label{fig:bivar}
\end{figure}

Sample variance is, of course, still evident, as is bias due to non-negativity of the GGC sample statistic (which may be countered by standard surrogate data methods), but both remain well below their minimum values across all model orders for the dual-regression case \citep[as evidenced by][Fig.~2]{StokesPurdon:2017}. \figref{fig:bivar} further compares bias and variance of time-domain GGC for the example system for single and dual regressions, at model order 3, across a wide-range of time-series lengths. A single regression consistently leads to substantially less bias and variance, except at high time-series lengths, where there is a drop off of bias and variance for both methods.

\cite{StokesPurdon:2017} do correctly identify a fundamental cause of the problem with dual-regression GGC estimation: even if the full process is a finite-order autoregression, the reduced process will generally \emph{not} be finite-order autoregressive; rather, it will be VARMA, or equivalently, a finite-order state-space process \citep{HandD:2012} -- which may be poorly modelled as a finite-order VAR \citep{Barnett:mvgc:2014}. The problem is in fact more pervasive than this: the full process \emph{itself} may have a strong moving-average (MA) component and be poorly-modelled as a finite-order VAR. This is because common features of neurophysiological data acquisition, sampling and preprocessing procedures such as subsampling and other temporal aggregation, filtering, measurement noise and sub-process extraction will all, in general, induce an MA component \citep{Barnett:gcfilt:2011,Seth:gcfmri:2013,Solo:2016}. This is particularly pertinent to fMRI data, where the haemodynamic response acts as a slow, MA filter. Fortunately, the state-space and non-parametric approaches mentioned above handle VARMA data parsimoniously, hence avoiding this problem.

Moving on to the second claim, Stokes and Purdon note that GGC reflects a combination of `transmitter' and `channel' dynamics, and is independent of `receiver' dynamics. Again, this independence has been previously identified, as a direct consequence of the invariance of GGC under certain affine transformations \citep{barrett:pre:2010,Barnett:gcfilt:2011}. But why should this independence matter? They suggest that it runs ``counter to intuitive notions of causality intended to explain observed effects'' since, according to them, ``neuroscientists seek to determine the mechanisms that produce `effects' within a neural system or circuit as a function of inputs or `causes' observed at other locations''.  In fact, this view resonates more strongly with approaches such as Dynamic Causal Modelling \citep[DCM;][]{friston:2003}---usually characterised as `effective connectivity'---which attempt to find the optimal mechanistic (circuit-level) description that explains observed data. GGC, on the other hand, models dependencies among observed responses and is therefore an example of (directed) `functional connectivity' \citep[see][for in-depth comparison]{Seth:jneuro:2015,FristonEtal:2013}. Essentially, the distinction is between making inferences about an underlying \emph{physical causal mechanism} \citep[DCM;][]{ValdesSosaEtal:2011} and making inferences about \emph{directed information flow} \citep[GGC;][]{Barnett:tegc:2009}. DCM is able to deliver evidence for circuit-level descriptions of neural mechanism from a limited repertoire of tightly-framed hypotheses, which must be independently motivated and validated \citep{StephanEtal:dcm:2010}; it is, in particular, unsuited to \emph{exploratory} analyses. GGC, on the other hand, is data-driven and ``data-agnostic'' (it makes few assumptions about the generative process, beyond that it be reasonably parsimoniously modelled as a linear stochastic system), and as such is well-suited to exploratory analyses. It delivers an information-theoretic interpretation of the neural process which is both amenable to statistical inference, and which also stands as an \emph{effect size} for directed information flow between components of the system \citep{BarrettBarnett:2013}. Both approaches address valid questions of interest to neuroscientific analysis.

Concluding, GGC represents a conceptually satisfying and statistically powerful method for (directed) functional connectivity analysis in neuroscience and neuroimaging. Currently available implementations [\eg, \cite{Barnett:mvgc:2012}] deal appropriately with issues of bias and variance associated with dual regression methods. However, a range of additional challenges remain in further developing this useful technique. These include issues of stationarity, linearity and exogenous influences, as noted by \cite{StokesPurdon:2017}, and in addition the influences of noise, sampling rates and temporal/spatial aggregation engendered by neural data acquisition \citep{Solo:2016,Barnett:downsample:2017}.

\section*{Acknowledgements}

ABB is funded by EPSRC grant EP/L005131/1. All authors are grateful to the Dr. Mortimer and Theresa Sackler Foundation, which supports the Sackler Centre for Consciousness Science.

\bibliographystyle{model2-names}
\bibliography{BarnettBarrettSeth_SPrejoinder}

\end{document}